\documentclass{article}

\usepackage{microtype}
\usepackage{graphicx}
\usepackage{subcaption}
\usepackage{booktabs} 
\usepackage{soul}

\usepackage[pagebackref=true]{hyperref}

\usepackage[accepted]{icml2024}

\usepackage{amsmath}
\usepackage{amssymb}
\usepackage{mathtools}
\usepackage{amsthm}
\usepackage{physics}
\usepackage{bm}
\usepackage{booktabs}

\usepackage[capitalize,noabbrev]{cleveref}

\theoremstyle{plain}

\theoremstyle{definition}

\theoremstyle{remark}

\usepackage[textsize=tiny]{todonotes}

\usepackage{multirow}
\usepackage{pifont}

\newcommand{\fnorm}[1]{\lVert {#1} \rVert_F}
\newcommand{\specnorm}[1]{\lVert {#1} \rVert_2}
\newcommand{\ours}{ReP}

\newcommand{\tabnum}[2]{{#1{\footnotesize±}{\scriptsize#2}}}

\newcommand{\besttabnum}[2]{{\textbf{#1}{\footnotesize\textbf{±}}{\scriptsize\textbf{#2}}}}

\newcommand{\set}[1]{\mathcal{#1}}

\icmltitlerunning{Mitigating Oversmoothing Through Reverse Process of GNNs}

\begin{document}

\twocolumn[
\icmltitle{Mitigating Oversmoothing
           Through Reverse Process of GNNs \\ for Heterophilic Graphs}



\icmlsetsymbol{equal}{*}

\begin{icmlauthorlist}
\icmlauthor{MoonJeong Park}{aigs}
\icmlauthor{Jaeseung Heo}{aigs}
\icmlauthor{Dongwoo Kim}{aigs,cse}
\end{icmlauthorlist}

\icmlaffiliation{aigs}{Graduate School of Artificial Intelligence, Pohang University of Science and Technology (POSTECH), Pohang, Republic of Korea}
\icmlaffiliation{cse}{Computer Science and Engineering, Pohang University of Science and Technology (POSTECH), Pohang, Republic of Korea}
\icmlcorrespondingauthor{Dongwoo Kim}{dongwoo.kim@postech.ac.kr}

\icmlkeywords{Machine Learning, ICML}

\vskip 0.3in
]



\printAffiliationsAndNotice{}  

\begin{abstract}

Graph Neural Network (GNN) resembles the diffusion process, leading to the over-smoothing of learned representations when stacking many layers. Hence, the reverse process of message passing can produce the distinguishable node representations by inverting the forward message propagation. The distinguishable representations can help us to better classify neighboring nodes with different labels, such as in heterophilic graphs.
In this work, we apply the design principle of the reverse process to the three variants of the GNNs.
Through the experiments on heterophilic graph data, where adjacent nodes need to have different representations for successful classification, we show that the reverse process significantly improves the prediction performance in many cases. Additional analysis reveals that the reverse mechanism can mitigate the over-smoothing over hundreds of layers.
Our code is available at \url{https://github.com/ml-postech/reverse-gnn}.
\end{abstract}
\section{Introduction}

Graph neural networks (GNNs) have emerged as an important tool for learning relational data. Earlier attempts aim to learn the node representations from graphs based on a message-passing mechanism. The message-passing neural network framework shows partial success with the homophilic graphs, where the nodes with the same labels are likely to be connected. When the heterophilic graphs, where node labels significantly differ from those of their neighbors, are considered, the models based on the homophilic assumption~\citep{mcpherson2001birds} often perform worse than the naive neural network architectures without considering the relationship between nodes~\citep{zhu2020beyond}.

\begin{figure*}[t]
    \centering
    \includegraphics[width=\linewidth]{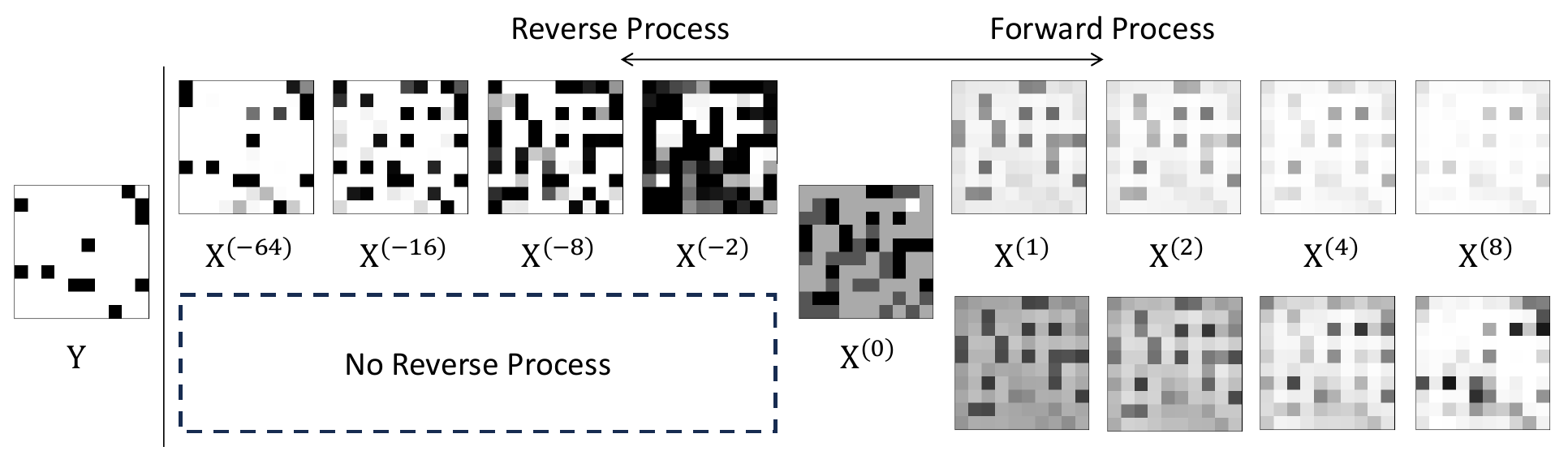}
    \caption{Visualized node representations over the forward and reverse processes in Minesweeper. Top: our approach with the forward and reverse processes. Bottom: a classical GCN with a forward process only.  The original node features are smoothed over the forward process, whereas the features are more distinguishable over the reverse process. Visualization details are provided in \cref{sec:mine-exp}.}
    \label{fig:minesweeper-2}
    \vspace{-3mm}
\end{figure*}

To learn the node representations of heterophilic graphs, a GNN needs to capture long-range interactions between nodes, leading to the stacking of multiple message-passing layers~\citep{li2022finding,rusch2023survey}. However, many studies empirically and theoretically identify that GNN tends to smooth the node representations over the layers, and eventually, the learned representations are likely to be similar, known as the over-smoothing issue~\citep{chen2020measuring,rusch2023survey}.  Furthermore, GRAND~\citep{pmlr-v139-chamberlain21a} shows that GNN can be seen as a discretization of a heat \emph{diffusion} equation. The diffusion perspective implies that learning distinguishable node representation with deep GNN is challenging since heat only diffuses to reach equilibrium, where node representation becomes indistinguishable. 


In this work, we claim not to forcefully correct the diffusive nature of the GNNs. Instead, we propose to use the reverse process of the aggregation. The aggregation process is known to make the node representations similar; hence, its reverse process can make the neighborhood representations more \emph{distinguishable}. Revisiting the diffusion perspective, applying the reverse process means learning the states in the past, which are further away from equilibrium and more distinguishable.

To illustrate our intuition, we showcase our experimental results on the Minesweeper dataset, a well-known heterophilic dataset~\citep{platonov2023critical}, in \cref{fig:minesweeper-2}. In Minesweeper, a board is a grid-structured graph where each node is initialized with the number of mines in the adjacent nodes $\mathbf{X}^{(0)}$, and the goal is to classify the location of mine $\mathbf{Y}$ correctly. 
The top row visualizes the learned node representations with our approach, and the bottom row visualizes the representations with the GCN~\citep{kipf2017semisupervised}.
With the forward-only method, such as GCN, the learned representations often fail to obtain distinguishable representations for classification. 
However, when the reverse process is applied to the initial features, we can obtain a distinguishable representation from the backward process.

To this end, we propose the framework of reverse process GNNs utilizing the inversion of forward message-passing layers. Specifically, we provide three variants of reverse process GNN based on three backbone models: 1) GRAND~\citep{pmlr-v139-chamberlain21a}, 2) GCN~\citep{kipf2017semisupervised}, and 3) GAT~\citep{veličković2018graph}. For GRAND, we directly use the numerical method to obtain the representations in a backward direction. For GCN and GAT, we adopt the idea of iResNet~\citep{pmlr-v97-behrmann19a} to make invertible message-passing layers that can model the reverse process.

The experimental results on heterophilic datasets show that the reverse process improves the prediction performance compared with the forward-only models.
Our investigation reveals that the reverse process produces distinguishable representations and enables the stacking of hundreds, even a thousand layers, mitigating over-smoothing.
Successfully stacking deep layers allows for the capture of long-range dependencies, which are crucial for performance on heterophilic datasets.
The experiments on homophilic datasets confirm that the reverse process does not harm the prediction performance when the aggregation mechanism is sufficient.

\section{Related Work}

Most studies on heterophilic data focus on identifying nodes with similar characteristics even among non-adjacent ones for aggregation.  GPR-GNN~\citep{chien2020adaptive} utilizes a trainable generalized PageRank for feature aggregation, learning important neighborhood ranges from the data and emphasizing information within those ranges for aggregation. CPGNN~\citep{zhu2021graph} introduces a learnable compatibility matrix to capture the information of non-adjacent homophilic nodes. FSGNN~\citep{maurya2022simplifying} proposes soft feature selection, wherein it adaptively selects neighbors to aggregate with different hop distances. GloGNN~\citep{li2022finding} employs a coefficient matrix that represents node-to-node relationships for aggregation, allowing the aggregation of information from all nodes. GBK-GNN~\citep{du2022gbk} proposes a bi-kernel graph neural network that separately handles homophilic and heterophilic nodes. It uses a selection gate to predict whether a node is homophilic or heterophilic and obtains features using the corresponding kernel based on the prediction. LRGNN~\citep{liang2023predicting} uses a low-rank approximation to compute a label relationship matrix, employing it for signed message passing.

However, aggregation still causes global node representations to become similar, known as over-smoothing, leading to performance degradation on heterophilic datasets.
To overcome this issue, H\textsubscript{2}GCN~\citep{zhu2020beyond} and Ordered GNN~\citep{song2023ordered} proposes to preserve non-aggregated representation separately. 
H\textsubscript{2}GCN learns node representation by separating ego-embedding and neighbor-embedding and employing intermediate representations.
Ordered GNN proposes ordering message passing to prevent the mixing of messages from different hops.

Several studies propose methods to adaptively learn appropriate filters that can handle various graph structures.
FAGCN~\citep{bo2021beyond} introduces a GNN framework with a self-gating mechanism to adaptively use low-frequency and high-frequency signals.
JacobiConv~\citep{wang2022powerful} uses Jacobi bases for spectral filter, whose orthogonality and flexibility enable adaptation to a wide range of graph signal densities. ACM-GCN~\citep{acm-gcn2022} utilizes a filterbank which combines low-pass and high-pass filters, and adaptively considers node-wise local information.


On the other hand, several studies tackle the over-smoothing issue. GRAND~\citep{pmlr-v139-chamberlain21a} enhances the understanding of over-smoothing from the perspective of the resemblance between GNN structures and the heat diffusion equation.
PairNorm~\citep{Zhao2020PairNorm:} proposes a normalization layer that remains the total pairwise feature distances constant.
DropEdge~\citep{Rong2020DropEdge:} randomly removes edges from the graph to cut off messages passing between adjacent nodes.
Half-Hop~\citep{halfhop_2023} up-sample edges and slow down the message passing.
While delaying or limiting over-smoothing can make node representations less smooth, learning difference-enhanced representations between adjacent nodes, which is helpful for heterophilic data, is still challenging.

\section{Method}
We aim to introduce a framework that can learn distinct node representations between adjacent nodes through reverse diffusion. We first provide the overall framework of our approach and then show three substantiations of the framework for three baseline models.

\subsection{Framework}
We consider a graph $\mathcal{G}=(\mathcal{V},\mathcal{E})$, where $\set{V}$ and $\set{E}$ are a set of nodes and edges respectively, with additional $d$-dimensional node features represented as $\mathbf{X}\in\mathbb{R}^{|\mathcal{V}|\times d}$ for all node, and $\mathbf{x}_i \in \mathbb{R}^{d}$ denotes the feature of node $i$.
It is a well-known fact that a typical GNN layer $f$ tends to learn similar representations between neighboring nodes, leading to the issue of over-smoothing.
\citet{pmlr-v139-chamberlain21a} highlighted that this is due to the diffusive property inherent in GNN structures.

In contrast to a typical GNN layer, we propose a GNN layer $g$ that performs the opposite role by reversing diffusion and re-concentrating diffused information.
Our main idea is to design an inverse function of a message-passing GNN layer $f$, with $g=f^{-1}$ to perform the reversion.
Due to the diffusive nature of the GNN layer $f$, its inverse form $g$ is expected to have two properties: 1) $g$ cancels the smoothing effect, producing distinguishable representations that mitigate the over-smoothing issue and thus 2) leading to stacking multiple layers of $g$.

Formally, with the inverse of multiple message passing layers, our framework predicts node label $\mathbf{Y}$ as follows:
\begin{equation*}\label{eq:label-prediction}
    \hat{\mathbf{Y}} = \phi(f^{(L)}\circ \cdots \circ f^{(1)}(\mathbf{X}^{(0)}) \Vert g^{(1)} \circ \cdots \circ g^{(L)}(\mathbf{X}^{(0)})),
\end{equation*}
where $\mathbf{X}^{(0)}$ is input node features, $f^{(\ell)}$ is the $\ell$-th forward message-passing layer with $g^{(\ell)} = f^{(\ell)^{-1}}$, $L$ is the number of layers, $\Vert$ denotes concatenation, and $\phi$ is a prediction function based on the forward and reverse processes of the input features.
We concatenate the representations from both directions for prediction to utilize 
advantage of difference enhanced representation and smooth representaiton at the same time. 
In practice, we can set the number of forward and reverse layers differently as $L_F$ and $L_R$, and share the parameters of different layers.

In the following sections, we propose a range of methods to develop a reverse diffusion function for GRAND and two variants of GNN with residual connections.

\subsection{Reverse Diffusion Based on GRAND}
In this section, we suggest a reverse diffusion function based on GRAND.
In GRAND, a GNN is interpreted as a discretization of the heat diffusion process.
This is modeled by the following heat diffusion equation on the graph:
\begin{equation}\label{eq:diff_process}
\frac{\partial \mathbf{X}(t)}{\partial t} =\big(\mathbf{A}(\mathbf{X}(t))-\mathbf{I}\big) \mathbf{X}(t),
\end{equation}
where ${\mathbf{A}}(\mathbf{X}) \in \mathbb{R}^{|\mathcal{V}|\times |\mathcal{V}|}$ represents the learnable attention matrix. 
Here, $[A(X)]_{ij}=0$ for any $(i,j)\notin\mathcal{E}$.

Within the framework of diffusion, the time parameter serves as a continuous layer, similar to the concept used in NeuralODEs \cite{neuralode}.
Using \cref{eq:diff_process}, GRAND produce node representations at time $T_F>0$ by:
\begin{equation}\label{eq:GRAND_rep}
    \mathbf{X}(T_F)=\mathbf{X}(0)+\int_{0}^{T_F} \pdv{\mathbf{X}(t)}{t}dt,
\end{equation}
where numerical techniques like the Euler method are used to solve integration.
Since \cref{eq:diff_process} models the property of heat reaching equilibrium over time, the node representations obtained through \cref{eq:GRAND_rep} become diffused as time progresses.
Conversely, tracing back in time allows us to observe the concentrated form of heat before diffusion.
Utilizing \cref{eq:diff_process}, node representations at a past time $T_R < 0$ can be calculated as follows:
\begin{equation}\label{eq:GRAND_past_emb}
\mathbf{X}(T_R)=\mathbf{X}(0)-\int_{T_R}^{0} \pdv{\mathbf{X}(t)}{t}dt.\;
\end{equation}
\cref{eq:GRAND_past_emb} reverses the diffusion process, enabling us to obtain distinguishable representations.

In experiments, we utilize the GRAND-l model, where the learnable attention matrix remains constant throughout the diffusion process, $\mathbf{A}(\mathbf{X}(t))=\mathbf{A}$, which is known to be parameter-efficient and robust to overfitting.
Following the original work, we use the scaled dot product attention to calculate the learnable attention matrix $\mathbf{A}(\mathbf{X})$, which is given as follows:
\begin{equation}
    [A(X)]_{ij}=\operatorname{softmax}\left(\frac{\left(\mathbf{W}_\text{K} \mathbf{x}_i\right)^{\top} \mathbf{W}_\text{Q} \mathbf{x}_j}{d'}\right),
    \;
\end{equation}
where $\mathbf{W}_\text{K}$ and $\mathbf{W}_\text{Q}$ are $d'\times d$ learnable parameters.
When multi-head attention is employed, we use the average of attentions, i.e., $\mathbf{A}(\mathbf{X})=\frac{1}{K}\sum_k \mathbf{A}^{(k)}(\mathbf{X})$, where $\mathbf{A}^{(k)}$ is the attention with $k$-th head.

\subsection{Reverse Process Based on GNN with Residual Connections}\label{sec:iresgnn}
We have explored the design of reverse process in widely used two message-passing GNN structures: graph convolutional network (GCN)~\citep{kipf2017semisupervised} and graph attention network (GAT)~\cite{veličković2018graph}. 
Both GCN and GAT model a node representation through an aggregation step, where neighborhood representations are combined, and an update step, where the aggregated representations are merged into the target node representation.  
Let $\hat{\mathbf{A}} \in \mathbb{R}^{|\mathcal{V}|\times|\mathcal{V}|}$ encode neighborhood structure in a graph, and $\mathbf{W}\in\mathbb{R}^{d\times d}$ be a matrix of learnable parameters. With the application of skip-connections \citep{he_resnet}, the GCN and GAT layers can be formalized as 
\begin{align} \label{eq:GNN-gen}
f(\mathbf{X}^{(\ell)}) = \mathbf{X}^{(\ell + 1)} &=\mathbf{X}^{(\ell)}+\sigma(\hat{\mathbf{A}}\mathbf{X}^{(\ell)}\mathbf{W})\\         &=\mathbf{X}^{(\ell)}+h(\mathbf{X}^{(\ell)}),\;
\end{align}
where $\sigma(\cdot)$ represents a non-linear activation function. $\hat{\mathbf{A}}$ is the renormalized adjacency matrix in GCN and a learnable attention matrix in GAT. 
We note that when using multi-head attention, we take averaging approach as in GRAND to keep invertibility.

According to \citet{pmlr-v97-behrmann19a}, the inverse of the GNN layer $g=f^{-1}$ exists if the $\text{Lip}(h)<1$, where $\text{Lip}(h)$ is Lipschitz constant of $h$.
With the contractive nonlinear activations like ReLU, ELU, and tanh, $\text{Lip}(h)<1$ is satisfied if 
\begin{equation}\label{eq:con}
    \sup_{\mathbf{X}\neq\mathbf{0}}\frac{\fnorm{\hat{\mathbf{A}}\mathbf{XW}}}{\fnorm{\mathbf{X}}} < 1\;,
\end{equation}
where $\fnorm{\cdot}$ denotes Frobenius norm.
When the condition is guaranteed, $g(\mathbf{X}^{(\ell)})$ can be computed via fixed point iteration as described in \cref{alg:res_inv}, resulting in $\mathbf{X}^{(\ell-1)}$. 

To ensure the invertibility of $f$ throughout the entire training procedure, we enforce the weight matrix $\mathbf{W}$ to satisfy the condition.
Since it is difficult to optimize the weight matrix while satisfying the condition, we normalize the weight matrix after each gradient descent step. Specifically, given that the left side of \cref{eq:con} is upper bounded by 
\begin{equation}
    \sup_{\mathbf{X}\neq\mathbf{0}}\frac{\fnorm{\hat{\mathbf{A}}\mathbf{XW}}}{\fnorm{\mathbf{X}}}
    \leq \specnorm{\hat{\mathbf{A}}}\fnorm{\mathbf{W}}\;,
\end{equation}
where $\specnorm{\cdot}$ denotes spectral norm, we use the upper bound to normalize the weight matrix.

Note that the spectral norm of $\hat{\mathbf{A}}$ is straightforward for the two models that we considered as baselines: GCN and GAT.
In GCN, $\hat{\mathbf{A}}=\tilde{\mathbf{D}}^{-\frac{1}{2}}\tilde{\mathbf{A}}\tilde{\mathbf{D}}^{-\frac{1}{2}}$ where $\tilde{\mathbf{A}}=\mathbf{A}+\mathbf{I}$ is the adjacency matrix $\mathbf{A}$ with added self-loops and $\tilde{\mathbf{D}}$ is the diagonal degree matrix of $\tilde{\mathbf{A}}$.
A spectral norm of renormalized adjacency matrix $\specnorm{\hat{\mathbf{A}}}=1$. In GAT,
 $\specnorm{\hat{\mathbf{A}}}=1$ since $\hat{\mathbf{A}}$ is right-stochastic.
Therefore, normalizing the weight matrix through its Frobenius norm is sufficient to guarantee the condition in \cref{eq:con}.
The upper bound normalization reduces the time complexity at the expense of the exact supremum calculation. 
In experiments, we find that the Frobenius upper bound can still result in a competitive performance.

\setlength{\textfloatsep}{10pt}
\begin{algorithm}[t]
   \caption{Inverse of GNN via fixed-point iteration}
   \label{alg:res_inv}
\begin{algorithmic}
    \STATE {\bfseries Input:} output of residual layer $\mathbf{X}^{(\ell)}$, residual block $h$, the number of fixed-point iterations $M$
    \STATE {\bfseries Output:} input of residual layer $\mathbf{X}^{(\ell-1)}$
    \STATE $\mathbf{X} \gets \mathbf{X}^{(\ell)}$
    \FOR{$m = 1, \ldots, M$}
        \STATE $\mathbf{X} \gets \mathbf{X} - h(\mathbf{X})$
    \ENDFOR
    \STATE {\bfseries Return} $\mathbf{X}$
\end{algorithmic}
\end{algorithm}

When the scaling coefficient $c<1$ is given, $\mathbf{W}$ is normalized to $\frac{c\mathbf{W}}{\fnorm{\mathbf{W}}}$ if $c<\fnorm{\mathbf{W}}$, in order to satisfy $\text{Lip}(h)<c$.
When multi-head attention with $K$ heads is employed for GAT, 
parameters of $k$-th head $\mathbf{W}^{(k)}$ is normalized to 
$\frac{c\mathbf{W}^{(k)}}{\frac{1}{K}\sum_{k=1}^K\fnorm{\mathbf{W}^{(k)}}}$ for all $k$,
since the upper bound result in $\frac{1}{K}\sum_{k=1}^K\fnorm{\mathbf{W}^{(k)}}$.
The derivation of the upper bound for multi-head attention is provided in \cref{sec:multihead-gat}.
Since residual block $h$ is an operator on a Banach space, and we constraint the $\text{Lip}(h)<1$, the convergence of \cref{alg:res_inv} is guaranteed by the Banach fixed point theorem \citep{pmlr-v97-behrmann19a}. Inversion error in practice are reported in \cref{sec:conv-anal}.

While the time complexity of a GCN mainly depends on the number of the forward layers $L_F$, the complexity of the reverse process depends on the number of fixed point iterations $M$ and of the number of reverse layers $L_R$. In our implementation, we run the fixed point iteration until convergence and backpropagate over the iterations. 
We provide the time and memory complexity analysis in \cref{tab:complexity}, and the proof is provided in \cref{apdx_CAsection}. 
An analysis on $M$ and the run-time with varying $L_R$ in real experiments are provided in \cref{exp:run}.

\begin{table*}
    \centering
    \begin{tabular}{lcc}
         \toprule &  GCN with Residual Connection& + Reverse Process\\
         \midrule Forward Time& $O(L_F|\mathcal{E}|d+L_F|\mathcal{V}|d^2)$ & $+\;O(M^2L_R|\mathcal{E}|d+ML_R|\mathcal{V}|d^2+M^2L_Rd^3)$\\
         Forward Memory& $O(|\mathcal{E}|+L_F|\mathcal{V}|d+d^2)$ & $+\;O(L_R|\mathcal{V}|d)$\\
         \midrule Backward Time & $O(L_F^2|\mathcal{E}|d+L_F|\mathcal{V}|d^2+L_F^2d^3)$ & $+\;O(M^2L_R^3|\mathcal{E}|d+ML_R^2|\mathcal{V}|d^2+M^2L_R^3d^3)$\\
          Backward Memory& $O(|\mathcal{E}|+|\mathcal{V}|d+d^2)$ & \ding{55} \\
         \bottomrule
    \end{tabular}
    \caption{Space and time complexity of GCN for the forward and reverse processes. We show an additional complexity when using reverse process, for simplicity. \ding{55} denotes there are no additional complexity. $L_F$ and $L_R$ represent the number of forward and reverse layers, respectively, $M$ is the number of fixed point iterations, and $d$ is the dimensionality of the node representation.}
    \label{tab:complexity}
    \vspace*{-3mm}
\end{table*}


\section{Experiments}
The experimental section focuses on validating two research questions: 1) Can the reverse process produce distinguishable representations? 2) Does the reverse process alleviate over-smoothing problems, enabling the construction of deeper layers?

Throughout this section, we denote models with additional reverse layers by \ours{} (\textbf{Re}verse \textbf{P}rocess).
For example, GCN$+$\ours{} indicates the GCN backbone with the reverse process.
We adopt weight sharing approach of GRAND-l for all experiments using \ours{}.

\subsection{Node Classification}\label{exp:main_perf}
In this section, we validate the effectiveness of our framework on node classification. 
Our primary focus is on assessing performance improvements in heterophilic datasets, while we have also evaluated performance on homophilic datasets.

\subsubsection{Datasets}
For the node classification task, we utilize a diverse set of datasets to assess our model.
For heterophilic data, we explore two Wikipedia graphs, Chameleon and Squirrel, and five additional datasets, Roman-Empire, Amazon-Ratings, Minesweeper, Tolokers, and Questions, introduced by \citet{platonov2023critical}.
We adopted the filtering process for Chameleon and Squirrel to prevent train-test data leakage as recommended by \citet{platonov2023critical}.
In the case of homophilic data, our selection includes three citation graphs:  Cora, CiteSeer, and PubMed, along with two Amazon co-purchase graphs, Computers and Photo.
The statistics of the datasets are summarized in \cref{apdx_datastat}.

\subsubsection{Experimental Setup and Baselines}

For the heterophilic datasets, we adopt the experimental setup from \citet{platonov2023critical}, which provides ten random train/validation/test splits.
We train a model with cross-entropy loss and report mean accuracy and standard deviation for multi-class classification datasets, including Chameleon, Squirrel, Roman-Empire, and Amazon-Ratings.
For binary classification datasets, including Minesweeper, Tolokers, and Questions, binary cross-entropy loss is used, and mean ROC-AUC and standard deviation are reported.

We benchmark several neural architectures as baselines, including classic GNN models like GCN \citep{kipf2017semisupervised}, GraphSAGE \citep{hamilton2017inductive}, GAT~\citep{veličković2018graph}, and Graph Transformer (GT) \citep{shi2020masked} for more complex attention mechanisms. These baselines are augmented with skip connections and layer normalization. In addition, modifications proposed in \citet{zhu2020beyond} are made to GAT and GT, resulting in GAT-sep and GT-sep models. For heterophily-specific models, we use 10 models including H\textsubscript{2}GCN~\citep{zhu2020beyond}, CPGNN~\citep{zhu2021graph}, GPR-GNN~\citep{chien2020adaptive}, FSGNN~\citep{maurya2022simplifying}, GloGNN~\citep{li2022finding}, FAGCN~\citep{bo2021beyond}, GBK-GNN~\citep{du2022gbk}, JacobiConv~\citep{wang2022powerful}, LRGNN~\citep{liang2023predicting}, Ordered GNN~\citep{song2023ordered}, ACM-GCN~\citep{acm-gcn2022}, and Dir-GNN~\citep{rossi2023edge}.

For the homophilic datasets, we adopt the experimental setup from \citet{he2021bernnet}, splitting datasets into 60\%/20\%/20\% train/validation/test sets and using ten random splits for averaging results.
We compare our framework against seven baselines: MLP, GCN~\citep{kipf2017semisupervised}, GAT~\citep{veličković2018graph}, APPNP~\citep{gasteiger2018predict}, ChebNet~\citep{defferrard2016convolutional}, GPR-GNN \citep{chien2020adaptive}, and BernNet~\citep{he2021bernnet}.

\paragraph{Validation} 
For all experiments, we set the number of epochs to 1,000 and apply early stopping when there is no performance improvement for 100 consecutive epochs.
For GRAND+\ours{}, we validate the hyperparameters that maximize the validation metric in the following ranges: learning rate~$\in [10^{-5}, 10^{-1}]$, $T_F, T_R \in [0,10]$, $d \in \{16, 32, 64, 128, 256, 512\}$, $K\in[1,8]$, $d' \in \{4, 8, 16, 32, 64, 128\}$. With the Euler or Runge-Kutta methods, we search the step size over $[0.5, 8]$ and tolerance scale over $[1,20000]$ with the Dormand-Prince method.
For GCN+\ours{} and GAT+\ours{}, we validate the hyperparameters in the following ranges: learning rate$\in [10^{-5}, 10^{-1}]$, the number of forward and reverse layers $L_F, L_R \in \{1,2,4,8,16,32,64,128,256,512\}$, dropout probability $\in [0, 0.9]$ with step size of $0.1$, $c \in \{0.1, 0.5, 0.9, 0.999, 0.99999\}$, convergence threshold for fixed point iteration $\in \{10^{-4}, 10^{-5}, 10^{-6}\}$, $d \in \{128, 256, 512, 1024, 2048\}$,   and $M \in \{8, 16, 32, 64\}$.
We fix the non-linear activation function to ReLU.

\subsubsection{Results}

\begin{table*}[t!]
    \centering
    \begin{tabular}{cccccccc} 

         \toprule & \multirow{2}{*}{Squirrel} & \multirow{2}{*}{Chameleon}& Roman- & Amazon- & \multirow{2}{*}{Minesweeper} & \multirow{2}{*}{Tolokers} & \multirow{2}{*}{Questions}\\
         & & & empire & ratings & & 
         \\
         \midrule
         SAGE* & \tabnum{36.09}{1.99} & \tabnum{37.77}{4.14} & \tabnum{85.74}{0.67} & \boldsymbol{\tabnum{53.63}{0.39}}& \tabnum{93.51}{0.57} & \tabnum{82.43}{0.44} & \tabnum{76.44}{0.62}\\
         GAT-sep* & \tabnum{35.46}{3.10} & \tabnum{39.26}{2.50} & \underline{\tabnum{88.75}{0.41}} & \tabnum{52.70}{0.62} & \tabnum{93.91}{0.35} & \tabnum{83.78}{0.43} & \tabnum{76.79}{0.71}\\
         GT* & \tabnum{36.30}{1.98} & \tabnum{38.87}{3.66} & \tabnum{86.51}{0.73} & \tabnum{51.17}{0.66} & \tabnum{91.85}{0.76} & \tabnum{83.23}{0.64} & \tabnum{77.95}{0.68}\\
         GT-sep* & \tabnum{36.66}{1.63} & \tabnum{40.31}{3.01} & \tabnum{87.32}{0.39} & \tabnum{52.18}{0.80} & \tabnum{92.29}{0.47} & \tabnum{82.52}{0.92} & \underline{\tabnum{78.05}{0.93}}\\
         \midrule
         H\textsubscript{2}GCN
         * & \tabnum{35.10}{1.15} & \tabnum{26.75}{3.64} & \tabnum{60.11}{0.52} & \tabnum{36.47}{0.23} & \tabnum{89.71}{0.31} & \tabnum{73.35}{1.01} & \tabnum{63.59}{1.46}\\ 
         CPGNN* & \tabnum{30.04}{2.03} & \tabnum{33.00}{3.15} & \tabnum{63.96}{0.62} & \tabnum{39.79}{0.77} & \tabnum{52.03}{5.46} & \tabnum{73.36}{1.01} & \tabnum{65.96}{1.95}\\ 
         GPR-GNN* & \tabnum{38.95}{1.99} & \tabnum{39.93}{3.30} & \tabnum{64.85}{0.27} & \tabnum{44.88}{0.34} & \tabnum{86.24}{0.61} & \tabnum{72.94}{0.97} & \tabnum{55.48}{0.91}\\
         FSGNN* & \tabnum{35.92}{1.32} & \tabnum{40.61}{2.97} & \tabnum{79.92}{0.56} & \tabnum{52.74}{0.83} & \tabnum{90.08}{0.70} & \tabnum{82.76}{0.61} & \boldsymbol{\tabnum{78.86}{0.92}}\\
         GloGNN* & \tabnum{35.11}{1.24} & \tabnum{25.90}{3.58} & \tabnum{59.63}{0.69} & \tabnum{36.89}{0.14} & \tabnum{51.08}{1.23} & \tabnum{73.39}{1.17} & \tabnum{65.74}{1.19}\\ 
         FAGCN* & \underline{\tabnum{41.08}{2.27}} & \tabnum{41.90}{2.72} & \tabnum{65.22}{0.56} & \tabnum{44.12}{0.30} & \tabnum{88.17}{0.73} & \tabnum{77.75}{1.05} & \tabnum{77.24}{1.26}\\
         GBK-GNN* & \tabnum{35.51}{1.65} & \tabnum{39.61}{2.60} & \tabnum{74.57}{0.47} & \tabnum{45.98}{0.71} & \tabnum{90.85}{0.58} & \tabnum{81.01}{0.67} & \tabnum{74.47}{0.86}\\
         JacobiConv* & \tabnum{29.71}{1.66} & \tabnum{39.00}{4.20} & \tabnum{71.14}{0.42} & \tabnum{43.55}{0.48} & \tabnum{89.66}{0.40} & \tabnum{68.66}{0.65} & \tabnum{73.88}{1.16}\\
         LRGNN & \tabnum{39.51}{2.12} & \tabnum{41.24}{2.95} & \tabnum{40.88}{1.84} & \tabnum{42.23}{4.85} & \tabnum{52.66}{6.40} & \tabnum{74.24}{1.37} & \tabnum{66.41}{1.75}\\
         Ordered GNN & \tabnum{38.96}{2.19} & \tabnum{38.04}{5.55} & \tabnum{80.12}{1.22} & \tabnum{49.66}{1.01} & \tabnum{90.21}{1.15} & \tabnum{81.42}{0.65} & \tabnum{73.36}{1.09}\\
         ACM-GCN & \tabnum{33.07}{3.03} & \tabnum{31.78}{3.35} & \tabnum{69.66}{0.62}\textsuperscript{\textdagger} & \tabnum{32.26}{2.06} & \tabnum{90.53}{0.56} & \tabnum{79.18}{0.77} & \tabnum{62.50}{4.05} \\
         Dir-GNN & \tabnum{40.39}{1.11} & \tabnum{41.26}{2.00} & \boldsymbol{\tabnum{91.23}{0.32}}\textsuperscript{\textdagger} & \tabnum{44.88}{0.84} & \tabnum{91.35}{0.65} & \tabnum{81.78}{0.83} & \tabnum{76.30}{0.99} \\
         \midrule
         GRAND& \tabnum{35.94}{1.64} &  \tabnum{37.71}{4.48} & \tabnum{75.19}{0.56} & \tabnum{49.34}{0.72} & \tabnum{90.41}{0.78} & \tabnum{78.38}{1.91} & \tabnum{76.22}{1.06}\\
         GRAND+\ours{}& \tabnum{40.75}{2.44} &  \tabnum{42.14}{3.62} & \tabnum{77.53}{0.62} & \tabnum{48.30}{0.60} & \tabnum{91.42}{0.78} & \tabnum{80.44}{1.64} & \tabnum{76.41}{1.04}\\ 
         $\Delta$& $+4.81\;(\uparrow)$ & $+4.43\;(\uparrow)$ & $+2.34\;(\uparrow)$ & $-1.04\;(\downarrow)$ & $+1.01\;(\uparrow)$ & $+2.06\;(\uparrow)$ & $+0.19\;(\uparrow)$\\  
         \midrule
         GAT* & \tabnum{35.62}{2.06} & \tabnum{39.21}{3.08} & \tabnum{80.87}{0.30} & \tabnum{49.09}{0.63} & \tabnum{92.01}{0.68} & \tabnum{83.70}{0.47} & \tabnum{77.43}{1.20}\\
         GAT+\ours{}& \tabnum{39.66}{2.00}  & \underline{\tabnum{43.24}{4.48}} & \tabnum{85.87}{0.64} & \tabnum{52.68}{0.27} & \underline{\tabnum{94.89}{0.33}} & \underline{\tabnum{84.52}{0.56}} & \tabnum{76.21}{0.74}\\ 
         & $(32/32)$ & $(64/32)$ & $(64/4)$ & $(16/2)$& $(32/16)$ & $(8/1)$ & $(64/1)$\\
         $\Delta$& $+4.04\;(\uparrow)$ & $+4.03\;(\uparrow)$ & $+5.00\;(\uparrow)$ & $+3.59\;(\uparrow)$ & $+2.11\;(\uparrow)$ & $+0.82\;(\uparrow)$ & $-1.22\;(\downarrow)$\\  
         \midrule
         GCN* & \tabnum{39.47}{1.47} & \tabnum{40.89}{4.12} & \tabnum{73.69}{0.74} & \tabnum{48.70}{0.63} & \tabnum{89.75}{0.52} & \tabnum{83.64}{0.67} & \tabnum{76.09}{1.27}\\
         GCN+\ours{}& \besttabnum{45.89}{1.45} & \besttabnum{47.57}{3.90} & \tabnum{86.43}{0.74} & \underline{\tabnum{52.75}{0.62}} & \besttabnum{96.05}{0.19} & \besttabnum{86.08}{0.84} & \tabnum{77.96}{0.96}\\
         & $(256/256)$ & $(128/256)$ & $(256/16)$ & $(32/1)$ & $(1/256)$ & $(128/64)$ & $(128/64)$\\
         $\Delta$& $+6.42\;(\uparrow)$ & $+6.68\;(\uparrow)$ & $+12.74\;(\uparrow)$ & $+4.05\;(\uparrow)$ & $+6.30\;(\uparrow)$ & $+2.44\;(\uparrow)$ & $+1.87\;(\uparrow)$\\ 
         \bottomrule
    \end{tabular}
    \caption{Test performance and standard deviation on \textit{heterophilic} datasets. $\Delta$ indicates the difference with and without \ours. We also report the number of forward and reverse layers below the performance of GCN$+$\ours{} and GAT$+$\ours{}. The best and the second-best are bolded and underlined, respectively.}
    \label{tab:hetero_main}
\end{table*}

\cref{tab:hetero_main} shows node classification results on the heterophilic datasets. 
Results marked with * and \textsuperscript{\textdagger} in \cref{tab:hetero_main} are obtained from \citet{platonov2023critical} and \citet{rossi2023edge}, and * in \cref{tab:homo_main} from \citet{he2021bernnet}.
Applying \ours{} shows performance improvement for all backbones across most heterophilic datasets, with the most significant and consistent improvement observed in GCN.
GCN+\ours{} achieves state-of-the-art performance in four out of seven datasets and the second-best performance in one.
In the datasets where GCN+\ours{} attained state-of-the-art performance, the number of reverse layers was consistently above 64.
This observation shows that deep layers of GNNs with \ours{} can achieve superior performance without over-smoothing.

\cref{tab:homo_main} shows node classification performance on the homophilic datasets. Due to spacing, we only report the results on three datasets. All results are reported in \cref{sec:homo-result}.
No significant performance changes were observed when \ours{} applied on homophilic node classification. The results confirm that the distinguishable representations do not harm the prediction performance for homophily datasets where the forward aggregation is sufficient.



\paragraph{Analysis on the Number of Forward and Reverse Layers} 
To investigate whether many layers of reverse process improve performance in heterophilic datasets and compare its effect with that of many forward layers, we trained GCN+\ours{} with varying pairs of steps on two datasets: Chameleon and Minesweeper.
Specifically, we vary the number of layers from 1 to 1024 in one direction. 

\begin{table}[t!]
    \centering
    \begin{tabular}{cccc} 
         \toprule & Cora & CiteSeer & PubMed \\ 
         \midrule
         MLP* & \tabnum{76.96}{0.95} & \tabnum{76.58}{0.88} & \tabnum{85.94}{0.22} \\
         vanilla-GCN* & \tabnum{87.14}{1.01} & \tabnum{79.86}{0.67} & \tabnum{86.74}{0.27} \\
         vanilla-GAT* & \tabnum{88.03}{0.79} & \besttabnum{80.52}{0.71} & \tabnum{87.04}{0.24}  \\
         APPNP* & \tabnum{88.14}{0.73} & \underline{\tabnum{80.47}{0.74}} & \tabnum{88.12}{0.31} \\
         ChevNet* & \tabnum{86.67}{0.82} & \tabnum{79.11}{0.75} & \tabnum{87.95}{0.28} \\
         GPR-GNN* & \besttabnum{88.57}{0.69} & \tabnum{80.12}{0.83} & \tabnum{88.46}{0.33} \\
         BernNet* & \underline{\tabnum{88.52}{0.95}} & \tabnum{80.09}{0.79} & \tabnum{88.48}{0.41} \\
         \midrule
         GRAND& \tabnum{85.53}{0.64} & \tabnum{74.95}{1.37} & \tabnum{88.81}{0.69} \\ 
         GRAND+\ours{}& \tabnum{85.73}{1.39} & \tabnum{75.78}{1.48} & \tabnum{89.03}{0.61} \\ 
         $\Delta$& $+0.20\;(\uparrow)$ & $+0.83\;(\uparrow)$ & $+0.22\;(\uparrow)$ \\ 
         \midrule
         GAT& \tabnum{87.67}{0.84} & \tabnum{77.36}{1.59} & \tabnum{89.66}{0.60} \\  
         GAT+\ours{}& \tabnum{87.93}{1.60}  & \tabnum{77.06}{1.60} & \underline{\tabnum{89.94}{0.61}} \\  
         & $(64/32)$ & $(32/128)$ & $(8/8)$ \\
         $\Delta$& $+0.26(\uparrow)$ & $-0.30(\downarrow)$ & $+0.28(\uparrow)$ \\   
         \midrule
         GCN& \tabnum{88.00}{1.42} & \tabnum{77.15}{1.44} & \tabnum{89.37}{0.52} \\ 
         GCN+\ours{}& \tabnum{87.63}{1.40}  & \tabnum{77.33}{1.65} & \besttabnum{89.96}{0.55} \\ 
         & $(32/512)$ & $(8/32)$ & $(32/32)$ \\
         $\Delta$& $-0.37(\downarrow)$ & $+0.18(\uparrow)$ & $+0.59(\uparrow)$ \\  
         \bottomrule
    \end{tabular}
    \caption{Test accuracy and standard deviation on \textit{homophilic} datasets. $\Delta$ indicates the difference with and without \ours. We also report the number of forward and reverse layers below the performance of GCN$+$\ours{} and GAT$+$\ours{}. The best and the second-best are bolded and underlined, respectively.}
    \label{tab:homo_main}
\end{table}

\begin{figure}[t!]
    \centering
    \includegraphics[width=\linewidth]{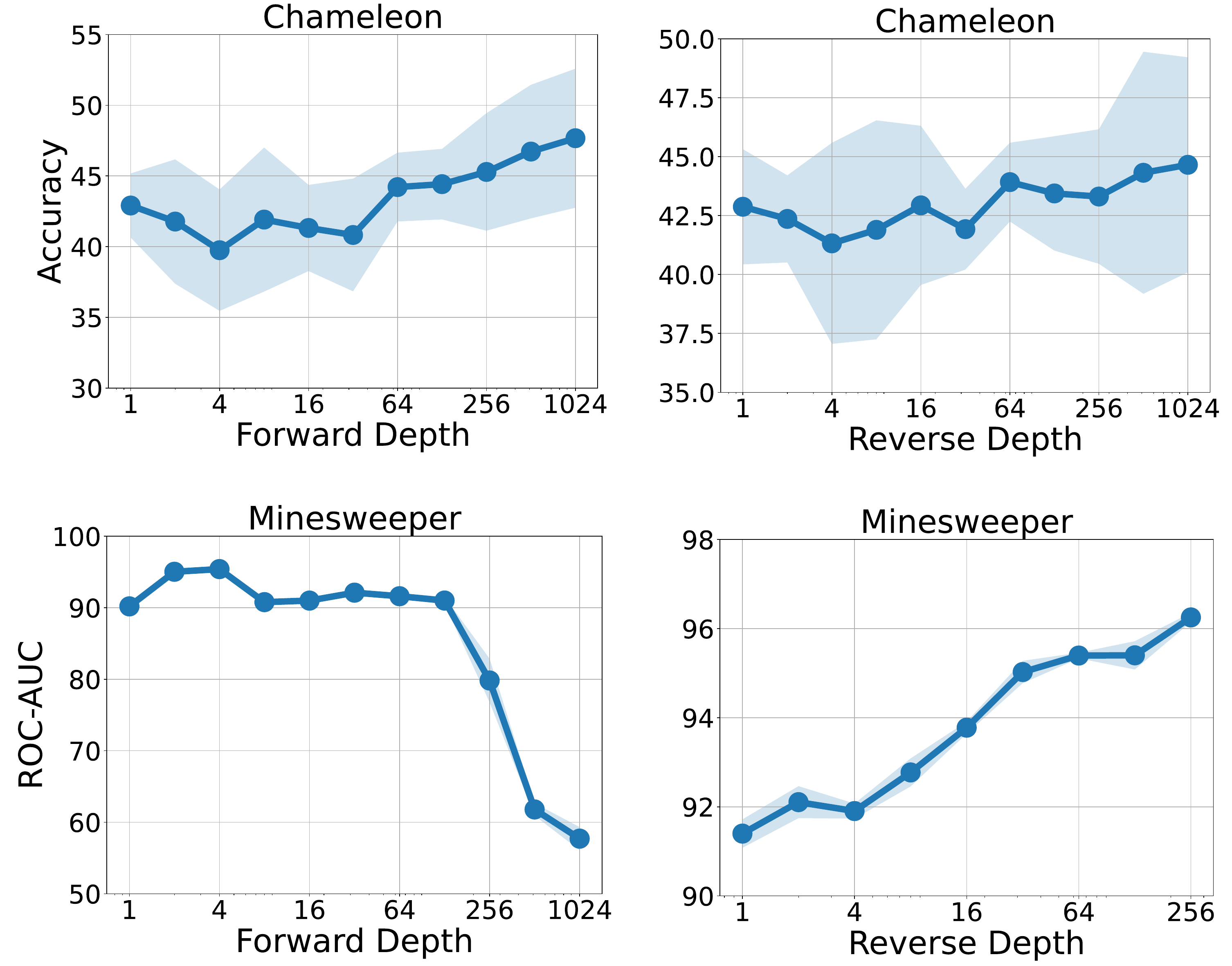}
    \caption{Prediction performance with varying the number of forward and reverse layers. We vary the number of layers (depth) in one direction. Due to memory constraints, we restricted the reverse depth used in Minesweeper to 256 or less.}
    \label{fig:iresgcn_depth}
\end{figure}

The prediction performances with varying numbers of layers are reported in \cref{fig:iresgcn_depth}.
In both datasets, the prediction performance keeps increasing as the number of reverse layers increases.
These results indicate that the reverse process is capable of deep stacking to mitigate over-smoothing. This enables the models to capture long-range dependencies effectively, which is known to be important, especially in heterophilic graphs.
The prediction performance also tends to increase as we increase the number of forward steps up to 1024 in the Chameleon dataset.

\paragraph{Over-Smoothing Analysis}\label{exp:deep_GSL}
We evaluate whether the proposed reverse process mitigates the over-smoothing issue. 
To measure the degree of over-smoothing, we adopt Graph Smoothness Level (GSL) proposed by \citet{kdd_hinders} defined as:
\begin{equation}
    \operatorname{GSL}(\mathbf{X}) = \frac{1}{|\mathcal{V}|(|\mathcal{V}|-1)} \sum_{i\in\mathcal{V}}\sum_{j \in \mathcal{V}, j \neq i} \frac{\mathbf{x}_i \cdot \mathbf{x}_j}{\fnorm{\mathbf{x}_i}\fnorm{\mathbf{x}_j}},\;
\end{equation}
where $\mathbf{x}_i$ is the representation of node $i$.
The GSL represents the average cosine similarity across all pairs of nodes in the graph.
A GSL value closer to one indicates more severe over-smoothing.

\begin{figure}
    \centering
    \includegraphics[width=\linewidth]{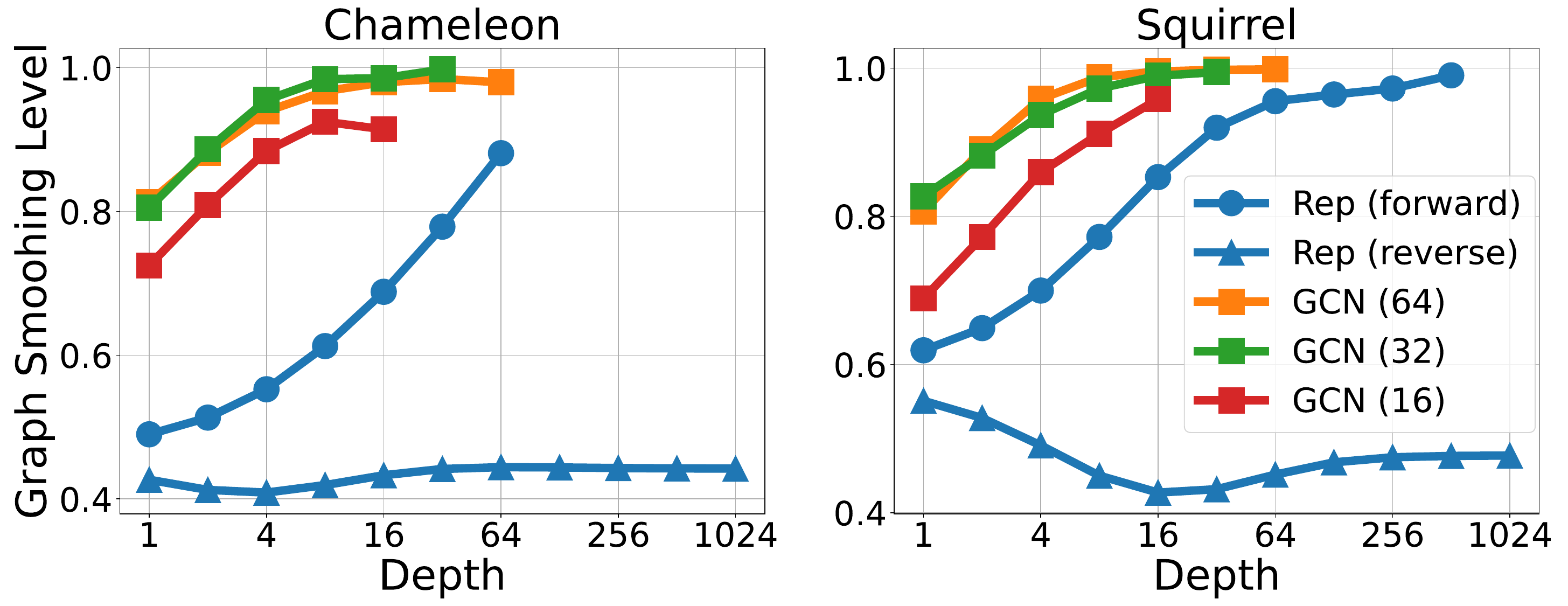}
    \caption{Over-smoothing levels measured by GSL over the number of layers (depth). \ours{} (forward) denotes the measured GSL in the forward process of GCN+\ours{}. We compare the results with GCN of three different depths: 16, 32, 64.}
    \label{fig:GSL}
\end{figure}

\cref{fig:GSL} shows GSL of GCN+\ours{} and GCN with varying numbers of layers on Squirrel and Chameleon datasets.
In both datasets, the GSL of GCN+\ours{} remains below 0.6 up to 1024 reverse layers, whereas the learned representations from GCN with 32 and 64 layers tend to become similar even after eight layers. 
GCN with 16 layers shows relatively low GSL values yet still exceeds 0.9 after eight layers.
Compared with CGN, GCN+\ours{} shows relatively less GSL, showing that the reverse process can mitigate the over-smoothing in the forward processes as well.

\paragraph{Qualitative Analysis on Minesweeper Dataset}\label{sec:mine-exp}
To validate that the reverse process produces a distinguishable representation, we visualize label predictions on the Minesweeper dataset.
The Minesweeper dataset is a binary classification task on a grid-structured graph, where the node with a positive label indicates the location of a mine. Each node, unless located on the boundaries, is connected to eight adjacent nodes, including the ones in the diagonal directions. The node feature is initialized with the number of mines in the adjacent nodes, and a one-hot representation of the feature is used as an initial representation of the node for learning.

\begin{figure}[t!]
    \centering
    \begin{subfigure}{0.48\columnwidth}
        \includegraphics[width=\linewidth]{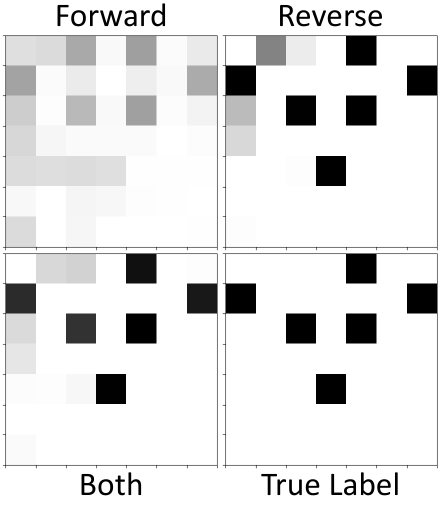}
        \subcaption{Example 1}
    \end{subfigure}
    \hfill
    \begin{subfigure}{0.48\columnwidth}
        \includegraphics[width=\linewidth]{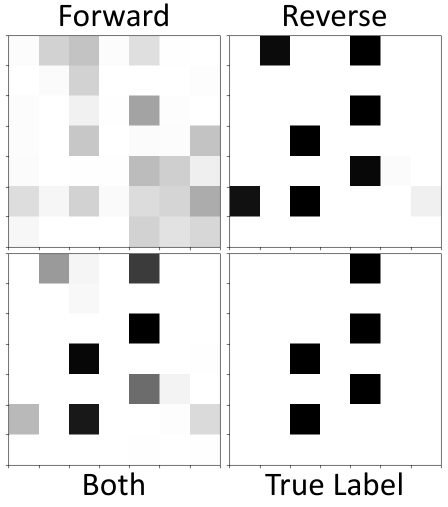}
        \subcaption{Example 2}
    \end{subfigure}
    \caption{Visualization of node prediction on the Minesweeper dataset. We visualize the prediction from 1) forward, 2) reverse, and 3) both representations, along with 4) the ground truth labels.}
    \label{fig:mine-qual}
\end{figure}

Based on the representations obtained from the forward and reverse processes, we trained a GCN+\ours{} model with a single-layer MLP as a prediction head.
In \cref{fig:mine-qual}, we visualize the prediction results of two randomly sampled $7 \times 7$ sub-grids from $100 \times 100$ grid structure. For each example, we visualize the prediction results with node representations from 1) forward process, 2) reverse process, and 3) both directions, as well as 4) the true labels, displayed from upper left to lower right.
In the visualization of the true labels, black cells indicate the presence of mine, and white cells indicate its absence.
In the visualization of the prediction results, the darker the cell, the higher the predicted probability of a mine being present.
Since the prediction head needs concatenated representations for prediction, to visualize the prediction results focused on a forward or reverse representation, we set the other node representation to be zero, e.g., to predict the mine using node representation at layer $\ell>0$, $(\mathbf{X}^{(\ell)}\Vert \mathbf{0})$ is fed into the prediction head.

In both examples, the prediction results from the reverse process appear distinguishable, while those from the forward process tend to be smooth.
Additionally, the distinguishable prediction created by the reverse process significantly contributes to label prediction.
Although in some cases, the reverse process can perfectly classify the location of mines, e.g., Example 1, the other cases require representations from both directions to classify the mines correctly, e.g., Example 2.

\cref{fig:minesweeper-2} shows the changes in predictions over the number of layers on the $10 \times 10$ sampled sub-grids. 
We follow the same visualization procedure with \cref{fig:mine-qual}.
As expected, the predictions and representations tend to be more distinguishable as the number of reverse layers increases.


\subsection{Inversion Error}
\label{sec:conv-anal}
Although the invertibility of \cref{alg:res_inv} is guaranteed in theory, the inversion may not be achieved due to numerical errors. To verify the fixed point method, we conduct the experiment to restore the inputs from the outputs of GCN and GAT at a depth of $64$. We set the scaling coefficient $c$ to $0.99999$ and the number of fixed point iterations to eight, which is challenging due to the large coefficient (note that the Lipschitz of 1 is non-invertible) and the small number of iterations. \cref{tab:inversion_error} shows the mean absolute error between the original inputs and restored input data. As the results show, the inversion error is negligible in practice.
\begin{table}[ht]
    \centering
    \begin{tabular}{ccc} 
         \toprule & GCN+\ours{} &  GAT+\ours{} \\
         \midrule
         Squirrel & $3.36\times10^{-5}$ & $3.28\times10^{-5}$ \\ 
         \midrule
         Chameleon & $2.23\times10^{-5}$ & $2.41\times10^{-5}$ \\
         \midrule
         Roman-empire & $2.79\times10^{-5}$ & $3.56\times10^{-5}$ \\
         \midrule
         Amazon-ratings & $4.31\times10^{-5}$ & $3.46\times10^{-5}$ \\
         \bottomrule
    \end{tabular}
    \caption{Mean absolute error between the original inputs and restored input data by \cref{alg:res_inv}.}
    \label{tab:inversion_error}
\end{table}

\subsection{Run Time Analysis}\label{exp:run}

\begin{figure}[t!]
    \centering
    \includegraphics[width=\linewidth]{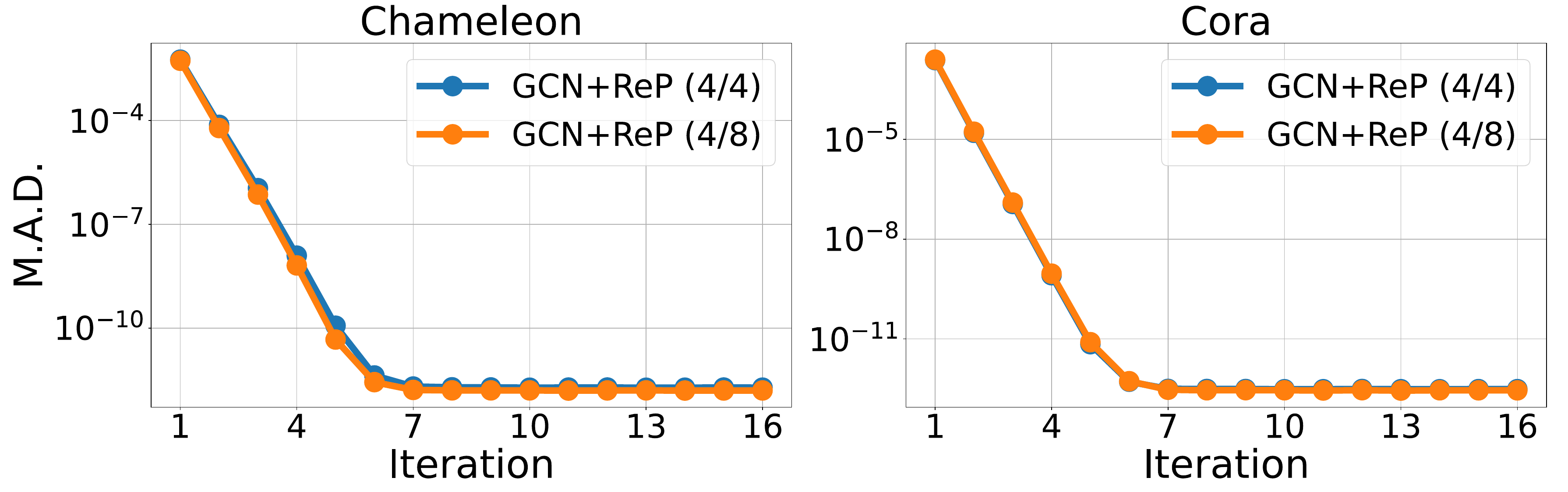}
    \caption{The mean absolute difference between two consecutive representations in the fixed-point method.}
    \label{fig:iter_diff}
\end{figure}

\begin{figure}[t!]
    \centering
    \includegraphics[width=\linewidth]{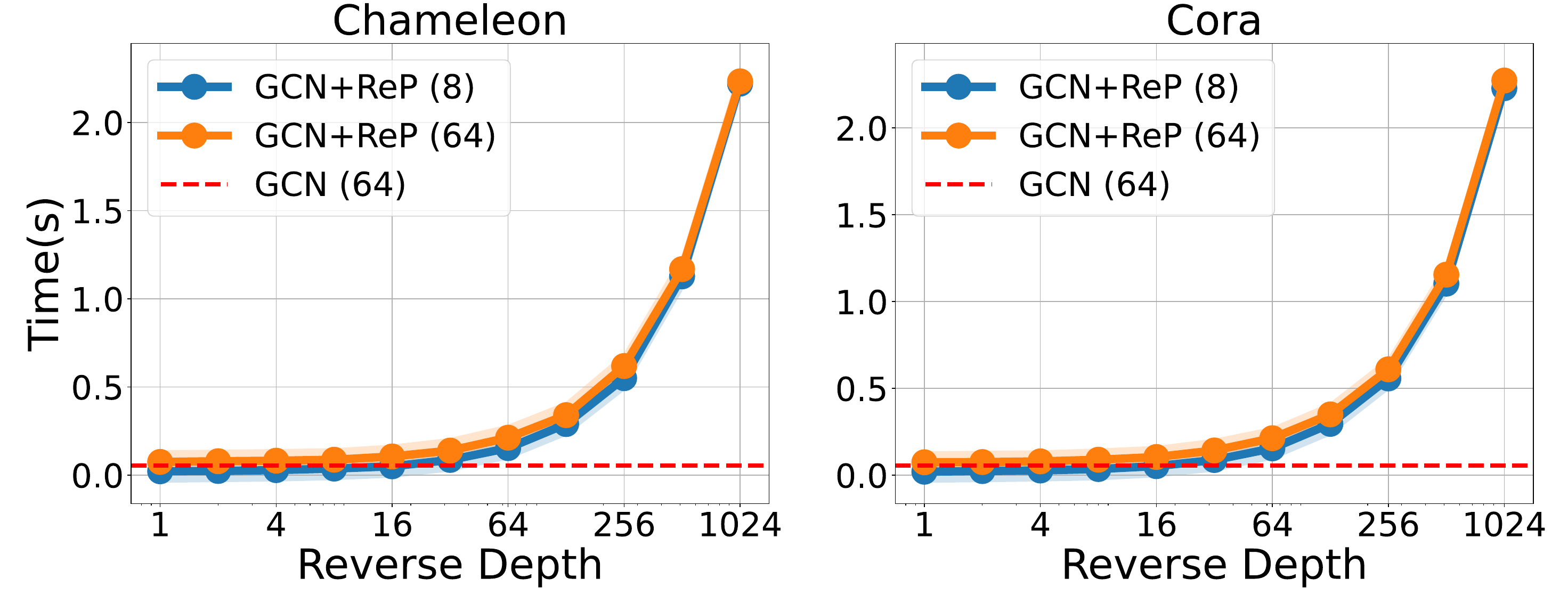}
    \caption{Average training time for the single epoch. The number of forward layers are shown next to the model.}
    \label{fig:runtime}
\end{figure}

We first measure how many iterations are required for the fixed point iterations to be converged.
\cref{fig:iter_diff} shows the difference between consecutive representations over the fixed point iterations with two datasets in terms of mean absolute difference. As shown in the figure, the fixed point method converges after seven iterations in general. 
In addition, we measure the training time for a single epoch and plot the results in 
\cref{fig:runtime}. The results show that the training time increases linearly as we increase the number of reverse layers coincided with the complexity analysis in \cref{sec:iresgnn}.


\section{Conclusion}
In this work, we propose a reverse process for the message-passing-based graph neural networks. Through extensive empirical analysis, we have found that the reverse process can mitigate over-smoothing issues and allow long-distance nodes to interact with each other. Especially for the heterophilic datasets where the long-range interaction is necessary for a better prediction, the proposed method achieves outstanding results against many baseline models.

\paragraph{Future Work} To ensure invertibility, the Lipschitz constant of the forward process must be restricted, and the hidden dimension of weight parameters must remain constant. These restrictions limit the representation power and design choices. Investigating less restrictive invertible forms could lead to performance improvements.


\section*{Impact Statement}
This paper presents work whose goal is to advance the field of Machine Learning. There are many potential societal consequences of our work, none of which we feel must be specifically highlighted here.

\section*{Acknowledgements}
This work was supported by Institute of Information \& communications Technology Planning \& Evaluation (IITP) grant funded by the Korea government(MSIT) (RS-2019-II191906, Artificial Intelligence Graduate School Program(POSTECH)) and supported by the National Research Foundation of Korea(NRF) grant funded by the Korea government(MSIT)(RS-2024-00337955 and RS-2023-00217286).

\bibliography{icml2024}
\bibliographystyle{icml2024}

\newpage
\appendix
\onecolumn



\section{Complexity Analysis}\label{apdx_CAsection}
The forward and reverse processes of a GCN with residual connections and weight sharing are as follows:
\begin{align}
    \mathbf{X}^{(\ell+1)}=\mathbf{X}^{(\ell)}+\hat{\mathbf{A}}\mathbf{X}^{(\ell)}\mathbf{W},&\quad\text{(forward process)}, \label{apdx_eq_diff_proc}\\
    \mathbf{X}^{(\ell-1)}=\mathbf{X}^{(\ell)}+\sum_{m=1}^M(-1)^m\hat{\mathbf{A}}^m\mathbf{X}^{(\ell)}\mathbf{W}^m,&\quad\text{(reverse process)}, \label{apdx_eq_inv_diff_proc}
\end{align}
where $M$ is the number of fixed point iterations.
For simplicity, we ignore an activation function. 

\subsection{Forward Pass of Forward Process}
\autoref{apdx_eq_diff_proc} involves three key operations: matrix multiplication between $\mathbf{X}$ and $\mathbf{W}$ with complexity $O(|\mathcal{V}|d^2)$, matrix multiplication between sparse matrix $\hat{\mathbf{A}}$ and $\mathbf{X}^{(\ell)}\mathbf{W}$ with complexity $O(|\mathcal{E}|d)$, and matrix addition with complexity $O(|\mathcal{V}|d)$. 
Overall, the time complexity for a single layer is $O(|\mathcal{V}|d^2+|\mathcal{E}|d)$. 
Therefore, the total time complexity over $L_F$ forward layers is $O(L_F|\mathcal{V}|d^2+L_F|\mathcal{E}|d)$.
Memory cost is calculated as $O(L_F|\mathcal{V}|d+|\mathcal{E}|+d^2)$.

\subsection{Backward Pass of Forward Process}
The time complexity of the backward pass is primarily determined by the computation cost of $\frac{\partial \mathcal{L}}{\partial \mathbf{W}}$.
By applying the chain rule, the gradient can be expressed as:
\begin{equation}
    \frac{\partial \mathcal{L}}{\partial \mathbf{W}} = \frac{\partial \mathcal{L}}{\partial \mathbf{X}^{(L_F)}} \frac{\partial \mathbf{X}^{(L_F)}}{\partial \mathbf{X}^{(L_F-1)}} \cdots \frac{\partial \mathbf{X}^{(2)}}{\partial \mathbf{X}^{(1)}} \frac{\partial \mathbf{X}^{(1)}}{\partial \mathbf{W}}.
\label{eq:backward_forward_basic}
\end{equation}

\newcommand{\lgrad}{\frac{\partial \mathcal{L}}{\partial \mathbf{X}^{(L_F)}}}
\newcommand{\X}{\mathbf{X}}
\newcommand{\A}{\hat{\mathbf{A}}}
\newcommand{\W}{\mathbf{W}}
\newcommand{\parfrac}[2]{\frac{\partial #1}{\partial #2}}
\newcommand{\XL}[1]{\X^{(L_F#1)}}

By sequentially multiplying $\parfrac{\X^{(n)}}{\X^{(n-1)}}$ for $n=L_F,\cdots,2$ on the right side of $\parfrac{\mathcal{L}}{\X^{(L_F)}}$, we derive:

\begin{equation}
    \frac{\partial \mathcal{L}}{\partial \mathbf{X}^{(L_F)}} \frac{\partial \mathbf{X}^{(L_F)}}{\partial \mathbf{X}^{(L_F-1)}} \cdots \frac{\partial \mathbf{X}^{(2)}}{\partial \mathbf{X}^{(1)}}
    = \sum_{\ell=0}^{L_F-1} \binom{L_F-1}{\ell} ({\hat{\mathbf{A}}^\top})^\ell \left( \frac{\partial \mathcal{L}}{\partial \mathbf{X}^{(L_F)}} \right) ({\mathbf{W}^\top})^\ell\;.
\label{eq:backward_forward_general}
\end{equation}

To demonstrate \cref{eq:backward_forward_general}, we illustrate part of the sequential multiplication process.
Multiplying $\parfrac{\X^{(L_F)}}{\X^{(L_F-1)}}$ to the right side of $\parfrac{\mathcal{L}}{\X^{(L_F)}}$, we get:

\begin{align}
    \parfrac{\mathcal{L}}{\X^{(L_F)}}\parfrac{\X^{(L_F)}}{\X^{(L_F-1)}}
    &=\parfrac{\mathcal{L}}{\X^{(L_F)}}
    \left( 
    \parfrac{}{\X^{(L_F-1)}}\mathbf{I}\X^{(L_F-1)}
    +\parfrac{}{\X^{(L_F-1)}}\A\X^{(L_F-1)}\W
    \right) \nonumber \\
    &=\parfrac{\mathcal{L}}{\X^{(L_F)}}+\A^\top \parfrac{\mathcal{L}}{\X^{(L_F)}} \W^\top.
\label{eq:backward_forward_first}
\end{align}

Next, we multiply $\parfrac{\X^{(L_F-1)}}{\X^{(L_F-2)}}$ on the right side of \cref{eq:backward_forward_first}, yielding:
\begin{align}
    \frac{\partial \mathcal{L}}{\partial \mathbf{X}^{(L_F)}} \frac{\partial \mathbf{X}^{(L_F)}}{\partial \mathbf{X}^{(L_F-1)}} \frac{\partial \mathbf{X}^{(L_F-1)}}{\partial \mathbf{X}^{(L_F-2)}} 
    &= \left(\parfrac{\mathcal{L}}{\X^{(L_F)}}+\A^\top \parfrac{\mathcal{L}}{\X^{(L_F)}} \W^\top\right)\left(\parfrac{}{\X^{(L_F-2)}}\mathbf{I}\X^{(L_F-2)}
    +\parfrac{}{\X^{(L_F-2)}}\A\X^{(L_F-2)}\W\right)\nonumber\\
    &= \sum_{\ell=0}^{2} \binom{2}{\ell} ({\hat{\mathbf{A}}^\top})^\ell \left( \frac{\partial \mathcal{L}}{\partial \mathbf{X}^{(L_F)}} \right) ({\mathbf{W}^\top})^\ell\;.
\end{align}

Repeating the process above, we can derive \cref{eq:backward_forward_general}.




Finally, $\frac{\partial \mathcal{L}}{\partial \mathbf{W}}$ is derived by multiplying \cref{eq:backward_forward_general} with $\parfrac{\X^{(1)}}{\W}$:
\begin{align}
    \frac{\partial \mathcal{L}}{\partial \mathbf{W}} 
    &= \left(\parfrac{\mathcal{L}}{\X^{(1)}}\right)\left(\parfrac{\X^{(1)}}{\W}\right) \nonumber \\
    &= \bigg(\sum_{\ell=0}^{L_F-1} \binom{L_F-1}{\ell} ({\hat{\mathbf{A}}^\top})^\ell \left( \frac{\partial \mathcal{L}}{\partial \mathbf{X}^{(L_F)}} \right) ({\mathbf{W}^\top})^\ell\bigg)\left(\frac{\partial }{\partial \mathbf{W}}\hat{\mathbf{A}}\mathbf{X}^{(0)}\mathbf{W}\right) \nonumber\\
    &=\left(\hat{\mathbf{A}}\mathbf{X}^{(0)}\right)^\top\bigg(\sum_{\ell=0}^{L_F-1} \binom{L_F-1}{\ell} ({\hat{\mathbf{A}}^\top})^\ell \left( \frac{\partial \mathcal{L}}{\partial \mathbf{X}^{(L_F)}} \right) ({\mathbf{W}^\top})^\ell\bigg).
\label{eq:backward_forward_final}
\end{align}
The time complexity of the term in the summation is $O(L_F+\ell|\mathcal{E}|d+|\mathcal{V}|d^2+\ell d^3)$. 
Therefore, the overall time complexity for computing $\frac{\partial \mathcal{L}}{\partial \mathbf{W}}$ is $O(L_F^2d^3+L_F|\mathcal{V}|d^2+L_F^2|\mathcal{E}|d)$, with a memory cost of $O(|\mathcal{V}|d+|\mathcal{E}|+d^2)$.

\subsection{Forward Pass of Reverse Process}
To calculate \autoref{apdx_eq_inv_diff_proc}, the initial step involves computing $(-1)^m\hat{\mathbf{A}}^m\mathbf{X}^{(\ell)}\mathbf{W}^m$. The time complexity for this part is $O(m|\mathcal{E}|d+|\mathcal{V}|d^2+md^3)$. Summing over $m$ from 0 to $M$, the overall complexity becomes $O(M^2|\mathcal{E}|d+M|\mathcal{V}|d^2+M^2d^3)$.
With $L_R$ representing the number of reverse process layers, additional time complexity becomes $O(M^2L_R|\mathcal{E}|d+ML_R|\mathcal{V}|d^2+M^2L_Rd^3)$ comparing to forward pass without reverse process.

\subsection{Backward Pass of Reverse Process}
The time complexity of the backward pass is primarily determined by the computation cost of $\frac{\partial \mathcal{L}}{\partial \mathbf{W}}$.
By applying the chain rule, the gradient can be expressed as:
\begin{equation}
    \frac{\partial \mathcal{L}}{\partial \mathbf{W}} = \frac{\partial \mathcal{L}}{\partial \mathbf{X}^{(-L_R)}} \frac{\partial \mathbf{X}^{(-L_R)}}{\partial \mathbf{X}^{(-L_R+1)}} \cdots \frac{\partial \mathbf{X}^{(-2)}}{\partial \mathbf{X}^{(-1)}} \frac{\partial \mathbf{X}^{(-1)}}{\partial \mathbf{W}}.
\end{equation}

By sequentially multiplying $\parfrac{\X^{(-n)}}{\X^{(-n+1)}}$ for $n=L_R,\cdots,2$ to the right side of $\parfrac{\mathcal{L}}{\X^{(-L_R)}}$, we derive:
\begin{align}
    &\frac{\partial \mathcal{L}}{\partial \mathbf{X}^{(-L_R)}} \frac{\partial \mathbf{X}^{(-L_R)}}{\partial \mathbf{X}^{(-L_R+1)}} \cdots \frac{\partial \mathbf{X}^{(-2)}}{\partial \mathbf{X}^{(-1)}}  \nonumber\\
    =& \frac{\partial \mathcal{L}}{\partial \mathbf{X}^{(-L_R)}}+\sum_{\ell=1}^{L_R-1}\binom{L_R-1}{\ell}\sum_{(m_1,m_2,\cdots,m_{\ell})}(-1)^{\sum_{i=1}^{\ell}m_i}({\hat{\mathbf{A}}^\top})^{\sum_{i=1}^{\ell}m_i}\left(  \frac{\partial \mathcal{L}}{\partial \mathbf{X}^{(-L_R)}} \right)({\mathbf{W}^\top})^{\sum_{i=1}^{\ell}m_i}\;,
\label{eq:backward_reverse_general}
\end{align}
where $m_i=1,\cdots,M$ for all $i$.
To demonstrate \cref{eq:backward_reverse_general}, we show part of the sequential multiplication process.
Multiplying $\parfrac{\X^{(-L_R)}}{\X^{(-L_R+1)}}$ to the right side of $\parfrac{\mathcal{L}}{\X^{(-L_R)}}$, we get:

\begin{align}
    \parfrac{\mathcal{L}}{\X^{(-L_R)}}
    \parfrac{\X^{(-L_R)}}{\X^{(-L_R+1)}}&=
    \parfrac{\mathcal{L}}{\X^{(-L_R)}}
    \parfrac{}{\X^{(-L_R+1)}}
    \left(
    \X^{(-L_R+1)}+\sum_{m=1}^M(-1)^m\A^m\X^{(-L_R+1)}\W^m
    \right) \nonumber\\
    &=\parfrac{\mathcal{L}}{\X^{(-L_R)}}+
    \sum_{m=1}^{M}(-1)^m\left(\A^m\right)^\top\parfrac{\mathcal{L}}{\X^{(-L_R)}}\left(\W^m\right)^\top\;,
\label{eq:backward_reverse_first}
\end{align}
which can also be obtained by substitute $L_R=2$ in \cref{eq:backward_reverse_general}.

Next, we multiply $\parfrac{\X^{(-L_R+1)}}{\X^{(-L_R+2)}}$ on the right side of \cref{eq:backward_reverse_first}, yielding:

\begin{align}
    &\frac{\partial \mathcal{L}}{\partial \mathbf{X}^{(-L_R)}} \frac{\partial \mathbf{X}^{(-L_R)}}{\partial \mathbf{X}^{(-L_R+1)}} \frac{\partial \mathbf{X}^{(-L_R+1)}}{\partial \mathbf{X}^{(-L_R+2)}}  \nonumber\\
    =&\left(\parfrac{\mathcal{L}}{\X^{(-L_R)}}+
    \sum_{m=1}^{M}(-1)^m\left(\A^m\right)^\top\parfrac{\mathcal{L}}{\X^{(-L_R)}}\left(\W^m\right)^\top\right)\parfrac{}{\X^{(-L_R+2)}}
    \left(
    \X^{(-L_R+2)}+\sum_{m=1}^M(-1)^m\A^m\X^{(-L_R+2)}\W^m
    \right)\nonumber\\
    =& \frac{\partial \mathcal{L}}{\partial \mathbf{X}^{(-L_R)}}
    +\sum_{\ell=1}^{2}\binom{2}{\ell}\sum_{(m_1,m_2,\cdots,m_{\ell})}(-1)^{\sum_{i=1}^{\ell}m_i}({\hat{\mathbf{A}}^\top})^{\sum_{i=1}^{\ell}m_i}\left(  \frac{\partial \mathcal{L}}{\partial \mathbf{X}^{(-L_R)}} \right)({\mathbf{W}^\top})^{\sum_{i=1}^{\ell}m_i}.
\end{align}
Repeating the process above, we can derive \cref{eq:backward_reverse_general}.
Finally, $\frac{\partial \mathcal{L}}{\partial \mathbf{W}}$ is derived by multiplying \cref{eq:backward_reverse_general} with $\parfrac{\X^{(-1)}}{\W}$:
\begin{align}
    \frac{\partial \mathcal{L}}{\partial \mathbf{W}} &= \sum_{m_0=1}^M(-1)^{m_0}m_0(\hat{\mathbf{A}}^{m_0}\mathbf{X}^{(0)})^\top\left(\frac{\partial \mathcal{L}}{\partial \mathbf{X}^{(-L_R)}}\right)\mathbf{W}^{m_0-1} \nonumber \\
    &+ \sum_{\ell=1}^{L_R-1}\binom{L_R-1}{\ell}\sum_{(m_0,m_1,\cdots,m_{\ell})}(-1)^{\sum_{i=1}^{\ell}m_i}m_0(\hat{\mathbf{A}}^{m_0}\mathbf{X}^{(0)})^\top({\hat{\mathbf{A}}^\top})^{\sum_{i=1}^{\ell}m_i} \left(  \frac{\partial \mathcal{L}}{\partial \mathbf{X}^{(-L_R)}} \right)({\mathbf{W}^\top})^{\sum_{i=1}^{\ell}m_i}\mathbf{W}^{m_0-1}\;,
\end{align}
where $m_i=1,\cdots,M$ for all $i$.
The time complexity of the first term is $O(M^2|\mathcal{E}|d+M|\mathcal{V}|d^2+M^2d^3)$. In the case of the second term, the time complexity is $O(M^2L_R^3|\mathcal{E}|d+ML_R^2|\mathcal{V}|d^2+M^2L_R^3d^3)$, since there are up to $2\ell M$ possible outcomes for $\sum_{i=1}^{\ell}m_i$. 
Therefore, the overall time complexity of the reverse process is $O((L_F^2+M^2L_R^3)|\mathcal{E}|d+(L_F^2+ML_R^2)|\mathcal{V}|d^2+M^2L_R^3d^3)$. 
The memory complexity is $O(|\mathcal{V}|d+|\mathcal{E}|+d^2)$.

\section{Dataset Statistics}\label{apdx_datastat}
\cref{tab:appendix_dataset_statistics} presents the dataset statistics utilized in experiments. There are two forms of homophily: edge homophily~\citep{abu2019mixhop, zhu2020beyond} and adjusted homophily~\citep{platonov2023characterizing}. Edge homophily denotes the proportion of edges connecting nodes with the same label, formally expressed as:
$$
h_{\text {edge }}=\frac{\left.\mid(u, v) \in E: y_u=y_v\right\} \mid}{|E|},
$$
where $E$ is the set of edges, and $y_n$ is the label of node $n$.
However, edge homophily is acknowledged to be meaningless in graphs with imbalanced labels. To address this issue, adjusted homophily is introduced. Formally, adjusted homophily is defined as:
$$
h_{a d j}=\frac{h_{\text {edge }}-\sum_{k=1}^C D_k^2 /(2|E|)^2}{1-\sum_{k=1}^C D_k^2 /(2|E|)^2},
$$
where $D_k$ is the total degree of nodes of class $k$, and $C$ is the number of classes. We employed seven heterophilic datasets characterized by low adjusted homophily and five commonly used homophilic datasets exhibiting high edge homophily and adjusted homophily.

\begin{table*}[h]
\begin{center}
\begin{tabular}{crrrrrrr}
\toprule
Dataset & \# nodes & \# edges & \# classes & avg degree &  edge homophily & adjusted homophily
\\ 
\midrule
Squirrel-filtered& 2,223 & 46,998 & 5 & 42.28 & 0.21 & 0.01\\
Chameleon-filtered& 890 & 8,854 & 5 & 19.90 & 0.24 & 0.03\\
roman-empire& 22,662 & 32,927 & 18 & 2.91 & 0.05 & -0.05 \\
amazon-ratings& 24,492 & 93,050 & 5 & 7.60 & 0.38 & 0.14 \\
minesweeper& 10,000 & 39,402 & 2 & 7.88 & 0.68 & 0.01 \\
tolokers& 11,758 & 519,000 & 2 & 88.28 & 0.59 & 0.09 \\
questions& 48,921 & 153,540 & 2 & 6.28 & 0.84 & 0.02 \\
\hline
Cora & 2,708 & 5,278 & 7 & 3.90 & 0.81 & 0.77 \\
CiteSeer & 3,327 & 4,552 & 6 & 2.74 & 0.74 & 0.67 \\
PubMed & 19,717 & 44,324 & 3 & 4.50 & 0.80 & 0.69 \\
Computers & 13,752 & 245,861 & 10 & 35.76 & 0.78 & 0.68 \\
Photo & 7,650 & 119,081 & 8 & 31.13 & 0.83 & 0.79 \\
\bottomrule
\end{tabular}
\caption{Statistics of the dataset utilized in the experiments.}
\label{tab:appendix_dataset_statistics}
\end{center}

\end{table*}

\section{Multi-Head Attention for GAT}
\label{sec:multihead-gat}
GAT calculates an attention matrix as follows:
\begin{equation}
    \hat{A}_{i j}=\frac{\exp \left(\operatorname{LeakyReLU}\left(\mathbf{a}^\top\left[\mathbf{W}^\top \mathbf{X}_i \| \mathbf{W}^\top \mathbf{X}_j\right]\right)\right)}{\sum_{k \in \mathcal{N}_i} \exp \left(\operatorname{LeakyReLU}\left(\mathbf{a}^\top\left[\mathbf{W}^\top \mathbf{X}_i \| \mathbf{W}^\top \mathbf{X}_j\right]\right)\right)},\;
\end{equation}
where $\mathbf{a}\in\mathbb{R}^{2d}$ is a learnable parameter.
Our framework also adopts averaging when using multi-head attention and remains hidden dimension constant to ensure invertibility, resulting in:
\begin{equation}
    h(\mathbf{X}^{(\ell)}) = \sigma(\frac{1}{K}\sum_{k=1}^K\hat{\mathbf{A}}^{(k)}\mathbf{X}^{(\ell)}\mathbf{W}^{(k)}).\;
\end{equation}
In this case, the Lipschitz constant of $h$, $\text{Lip}(h)<1$ is satisfied if
\begin{equation}
    \sup_{\mathbf{X}\neq\mathbf{0}}\frac{\fnorm{\frac{1}{K}\sum_{k=1}^K\hat{\mathbf{A}}^{(k)}\mathbf{X}^{(\ell)}\mathbf{W}^{(k)}}}{\fnorm{\mathbf{X}}} < 1\;.
\end{equation}
The upper bound of left side is computed by:
\begin{align}
    \sup_{\mathbf{X}\neq\mathbf{0}}\frac{\fnorm{\frac{1}{K}\sum_{k=1}^K\hat{\mathbf{A}}^{(k)}\mathbf{X}^{(\ell)}\mathbf{W}^{(k)}}}{\fnorm{\mathbf{X}}} 
    &\leq \frac{1}{K}\sum_{k=1}^K\sup_{\mathbf{X}\neq\mathbf{0}}\frac{\fnorm{\hat{\mathbf{A}}^{(k)}\mathbf{X}^{(\ell)}}}{\fnorm{\mathbf{X}}}\fnorm{\mathbf{W}^{(k)}} \nonumber\\
    &\leq \frac{1}{K}\sum_{k=1}^K \fnorm{\mathbf{W}^{(k)}},\;
\end{align}
since $\fnorm{\sum \mathbf{X}}\leq\sum\fnorm{\mathbf{X}}$.

\section{Full Results of Homophily Datasets}
\label{sec:homo-result}

We provide the experimental results of all homophilic datasets in \cref{tab:homo_appendix}.

\begin{table}[t!]
    \centering
    \begin{tabular}{cccccc} 
         \toprule & Cora & CiteSeer & PubMed & Computers & Photo \\ 
         \midrule
         MLP* & \tabnum{76.96}{0.95} & \tabnum{76.58}{0.88} & \tabnum{85.94}{0.22} & \tabnum{82.85}{0.38} & \tabnum{84.72}{0.34} \\
         vanilla-GCN* & \tabnum{87.14}{1.01} & \tabnum{79.86}{0.67} & \tabnum{86.74}{0.27} & \tabnum{83.32}{0.33} & \tabnum{88.26}{0.73} \\
         vanilla-GAT* & \tabnum{88.03}{0.79} & \besttabnum{80.52}{0.71} & \tabnum{87.04}{0.24} & \tabnum{83.32}{0.39} & \tabnum{90.94}{0.68} \\
         APPNP* & \tabnum{88.14}{0.73} & \underline{\tabnum{80.47}{0.74}} & \tabnum{88.12}{0.31} & \tabnum{85.32}{0.37} & \tabnum{88.51}{0.31} \\
         ChevNet* & \tabnum{86.67}{0.82} & \tabnum{79.11}{0.75} & \tabnum{87.95}{0.28} & \tabnum{87.54}{0.43} & \tabnum{93.77}{0.32} \\
         GPR-GNN* & \besttabnum{88.57}{0.69} & \tabnum{80.12}{0.83} & \tabnum{88.46}{0.33} & \tabnum{86.85}{0.25} & \tabnum{93.85}{0.28} \\
         BernNet* & \underline{\tabnum{88.52}{0.95}} & \tabnum{80.09}{0.79} & \tabnum{88.48}{0.41} & \tabnum{87.64}{0.44} & \tabnum{93.63}{0.35} \\
         \midrule
         GRAND& \tabnum{85.53}{0.64} & \tabnum{74.95}{1.37} & \tabnum{88.81}{0.69} & \tabnum{90.28}{0.47} & \tabnum{94.01}{0.73}\\ 
         GRAND+\ours{}& \tabnum{85.73}{1.39} & \tabnum{75.78}{1.48} & \tabnum{89.03}{0.61} & \tabnum{89.51}{0.78}  & \tabnum{94.48}{0.61} \\ 
         $\Delta$& $+0.20\;(\uparrow)$ & $+0.83\;(\uparrow)$ & $+0.22\;(\uparrow)$ & $-0.77\;(\downarrow)$ & $+0.47\;(\uparrow)$ \\ 
         \midrule
         GAT& \tabnum{87.67}{0.84} & \tabnum{77.36}{1.59} & \tabnum{89.66}{0.60} & \besttabnum{92.15}{0.30} & \besttabnum{95.86}{0.58}\\  
         GAT+\ours{}& \tabnum{87.93}{1.60}  & \tabnum{77.06}{1.60} & \underline{\tabnum{89.94}{0.61}} &\tabnum{91.03}{0.62} &\tabnum{95.44}{0.71} \\  
         & $(64/32)$ & $(32/128)$ & $(8/8)$ & $(16/8)$ & $(32/8)$\\
         $\Delta$& $+0.26(\uparrow)$ & $-0.30(\downarrow)$ & $+0.28(\uparrow)$ & $-1.12(\downarrow)$ & $-0.42(\downarrow)$\\   
         \midrule
         GCN& \tabnum{88.00}{1.42} & \tabnum{77.15}{1.44} & \tabnum{89.37}{0.52} & \underline{\tabnum{91.87}{0.57}} & \tabnum{95.35}{0.47}\\ 
         GCN+\ours{}& \tabnum{87.63}{1.40}  & \tabnum{77.33}{1.65} & \besttabnum{89.96}{0.55} & \tabnum{90.92}{0.52} & \underline{\tabnum{95.50}{0.63}}\\ 
         & $(32/512)$ & $(8/32)$ & $(32/32)$ & $(32/64)$ & $(32/128)$\\
         $\Delta$& $-0.37(\downarrow)$ & $+0.18(\uparrow)$ & $+0.59(\uparrow)$ & $-0.95(\downarrow)$ & $+0.15(\uparrow)$\\  
         \bottomrule
    \end{tabular}
    \caption{Test accuracy and standard deviation on \textit{homophilic} datasets. $\Delta$ indicates the difference with and without \ours. We also report the number of forward and reverse layers below the performance of GCN$+$\ours{} and GAT$+$\ours{}. The best and the second-best are bolded and underlined, respectively.}
    \label{tab:homo_appendix}
\end{table}



\end{document}